\documentclass[aps,twocolumn,showpacs,preprintnumbers,amsmath,amssymb,superscriptaddress, prb,aps]{revtex4}

\usepackage{graphicx}% Include figure files
\usepackage{epsfig}
\usepackage{dcolumn}% Align table columns on decimal point
\usepackage{bm}% bold math
\usepackage{longtable}
\usepackage{color}

\begin{document}

\title{Static magnetic moments revealed by muon spin relaxation and thermodynamic measurements in quantum spin ice Yb$_2$Ti$_2$O$_7$}

\author{Lieh-Jeng Chang}
\affiliation{Department of Physics, National Cheng Kung University, Tainan 70101, Taiwan}

\author{Martin R. Lees}
\affiliation{Department of Physics, University of Warwick, Coventry CV4 7AL, United Kingdom}

\author{Isao Watanabe}
\affiliation{Advanced Meson Science Laboratory, RIKEN Nishina Center, RIKEN, Wako 351-0198, Japan}

\author{Adrian D. Hillier}
\affiliation{ISIS Facility, Rutherford Appleton Laboratory, Chilton, Oxfordshire OX11 0QX, United Kingdom}

\author{Yukio Yasui}
\affiliation{Department of Physics, Meiji University, Kawasaki 214-8571, Japan}

\author{Shigeki Onoda}
\affiliation{Condensed Matter Theory Laboratory, RIKEN, Wako 351-0198, Japan}
\affiliation{Quantum Matter Theory Research Team, Center for Emergent Matter Science (CEMS), RIKEN, Wako 351-0198, Japan}

\begin{abstract}
We present muon spin relaxation ($\mu$SR) and specific-heat versus temperature $C(T)$ measurements on polycrystalline and single-crystal samples of the pyrochlore magnet Yb$_2$Ti$_2$O$_7$. $C(T)$ exhibits a sharp peak at a $T_\mathrm{C}$ of 0.21 and 0.26~K for the single-crystal and polycrystalline samples respectively. For both samples, the magnetic entropy released between 50~mK and 30~K amounts to $R\ln2$ per Yb. At temperatures below $T_\mathrm{C}$ we observe a steep drop in the asymmetry of the zero-field $\mu$SR time spectra at short time scales, as well as a decoupling of the muon spins from the internal field in longitudinal magnetic fields of $\leq0.25$~T for both the polycrystalline and single-crystal samples. These muon data are indicative of static magnetic moments. Our results are consistent with the onset of long-range magnetic order in both forms of Yb$_2$Ti$_2$O$_7$.
\end{abstract}


\maketitle
\section{INTRODUCTION}
Quantum spin ice has highlighted the quantum dynamics of emergent magnetic monopoles and the associated gauge fields~\cite{Hermele:04} in frustrated pyrochlore magnets.\cite{Gardner:10} In particular, the ground state of Yb$_2$Ti$_2$O$_7$ may be described as a U(1) quantum spin liquid hosting gapped bosonic spinon excitations carrying the monopole and fictitious gapless photon excitations~\cite{Hermele:04} or a ferromagnetically ordered Higgs phase achieved by a Bose condensation of spinons.\cite{Chang:12} In the latter case, the U(1) gauge invariance associated with the global phase of spinon wavefunctions is spontaneously broken and the gauge fields and thus the fictitious photons acquire a mass through the Anderson-Higgs mechanism.\cite{Anderson:63,Higgs:64} This provides a magnetic analogue of superconductivity on the lattice problem as an ideal laboratory for addressing the lattice gauge theory~\cite{Fradkin:79} in the U(1) case. In the early study, a first-order phase transition was observed around $T_C\sim0.24$~K on powder samples of Yb$_2$Ti$_2$O$_7$.\cite{Blote:69} However, the nature of the transition and the low-temperature phase have been 
the subject of intense debate.\cite{Hodges:02,Yasui:03,Chang:12,Ross:11,Ross:12,Yaouanc:11,DOrtenzio:13}

An emergence of ferromagnetic (FM) order has been reported for a single-crystal sample studied by neutron diffraction measurements.\cite{Yasui:03} Polarized neutron-scattering experiments on the same single crystal have shown a full depolarization of the incident neutron spins indicating the formation of ferromagnetic domains.\cite{Chang:12} The paramagnetic state is well described with a magnetic Coulomb liquid behavior characterized by the growth of a broadened pinch-point singularity~\cite{Henley:05} in the energy-integrated diffuse scattering profile with decreasing temperature,\cite{Chang:12} and the observed ordered structure well below the transition temperature $T_C\sim0.21$~K~\cite{Yasui:03} is consistent with the theoretical prediction of a formation of nearly collinear ferromagnetic moments parallel to $[100]$ in the Higgs phase.\cite{Chang:12,Ross:11}
% This ordered structure involves not only the $\langle111\rangle$ Ising component but also the planar component of each Yb magnetic moment. Since the planar component is described as the Bose condensation of spinons coupled to the U(1) gauge field within the framework of a quantum spin ice model~\cite{Onoda:10,Ross:11,Chang:12} and the transition temperature is low enough for a quantum Coulomb liquid behavior to appear~\cite{Banerjee:08}, this first-order phase transition has been regarded as a magnetic analogue of a Higgs transition~\cite{Chang:12}.

%In contrast to the crystal exhibiting FM order, in which there is a sharp anomaly at 0.21~K in temperature dependence of the specific heat $C(T)$, these other crystals are reported to have either a broader peak or no peak at all in $C(T)$~\cite{Chang:12,Ross:11,Yaouanc:11,Ross:12,DOrtenzio:13}.
%It is noteworthy that apart from the crystal studied in Refs.~\onlinecite{Yasui:03} and \onlinecite{Chang:12}, 
Long-range magnetic order, however, has only been detected in the single crystals of Yb$_2$Ti$_2$O$_7$ studied in Refs.~\onlinecite{Chang:12} and~\onlinecite{Yasui:03}. There is a sharp anomaly in the temperature dependence of the specific heat, $C(T)$ at $T_\mathrm{C}$ in these crystals, while other single crystal samples have either a broader peak or no peak at all in $C(T)$.\cite{Chang:12,Ross:11,Yaouanc:11,Ross:12,DOrtenzio:13} A possible crystallographic and/or chemical disorder of the crystals including the effects of strain and stuffing, i.e. an excess of Yb ions occupying the Ti sites, that has been detected from EXAFS~\cite{Chang:12} and neutron-scattering studies,\cite{Ross:12} may explain this discrepancy. 

Given these contrasting results on single crystals, experiments 
on polycrystalline samples may provide key information to solve this problem.
% although these data are also somewhat controversial.
Most authors report a sharp peak in $C(T)$ at low-temperature for powder samples of Yb$_2$Ti$_2$O$_7$,\cite{Blote:69,Yaouanc:11,Ross:12, DOrtenzio:13} while the peak temperature ranges from 0.24 to 0.265~K (Fig.~\ref{fig:C}(a)).
%Nevertheless, the behavior still varies from sample to sample. For instance, the peak in the specific heat appears at temperatures ranging from 0.24 to 0.265~K~\cite{Ross:12} and the width of the peak also varies. 
Muon spin relaxation ($\mu$SR) experiments on different powder samples have provided contradictory data: A short-time decay of the $\mu$SR asymmetry was observed at 0.2~K in an early study~\cite{Hodges:02} indicating at least a near freezing of the magnetic moments, while this behavior was not observed in more recent work.\cite{DOrtenzio:13} Moreover, long-range magnetic order has not been observed by neutron diffraction~\cite{Hodges:02} and spin echo~\cite{Gardner:04} experiments on polycrystalline Yb$_2$Ti$_2$O$_7$,
% Hodges \textit{et al}. observed no magnetic Bragg peaks in their powder neutron-diffraction data~\cite{Hodges:02}, 
although magnetic Bragg scattering at reciprocal lattice vectors are much harder to detect on powders than in single crystals. 
%Furthermore, the neutron spin depolarization observed on powders is not perfect~\cite{Gardner:04}; the neutron spin flipping ratio is 4, which is larger than the ideal value of 1 observed in the FM ordered single crystal (see Fig.~\ref{fig:C}(b))~\cite{Chang:12}. 
Comprehensive investigations on both powder and single-crystal Yb$_2$Ti$_2$O$_7$ are therefore required in order to resolve these controversies and to determine whether the ground state of quantum spin ice Yb$_2$Ti$_2$O$_7$ belongs to a magnetic Higgs phase or a U(1) quantum spin liquid.

In this paper, we present $\mu$SR and specific-heat measurements on both polycrystalline and high-quality single-crystal samples of Yb$_2$Ti$_2$O$_7$.  A sharp maximum in the temperature dependence of the specific heat reveals the presence of a magnetic transition at a $T_\mathrm{C}$ of 0.26 and 0.21~K for the powder and the single-crystal samples respectively, with a magnetic entropy released between 50~mK and 30~K that amounts to $R\ln2$ per Yb. Zero-field (ZF) and longitudinal-field (LF) $\mu$SR measurements show that below $T_\mathrm{C}$, static magnetic moments exist in both samples. The magnetic volume fractions are estimated to be $100\%$ and $80\%$ for the powder and single crystal, respectively. Our results are consistent with the onset of a magnetic order in Yb$_2$Ti$_2$O$_7$ in both forms of this material.

\section{EXPERIMENTAL DETAILS}

\subsection{SAMPLE PREPARATION AND HEAT CAPACITY MEASUREMENTS}
\begin{figure}[tb]
  \begin{center}\leavevmode
 \includegraphics[width=\columnwidth]{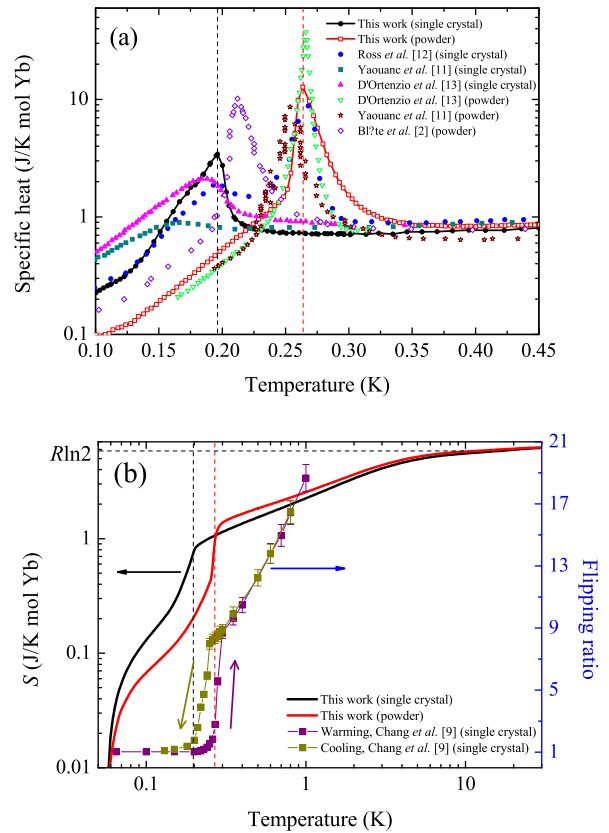}
  \end{center}	
  \caption{(Color online). (a) Specific heat versus temperature $C(T)$ for our single crystal~\cite{Chang:12} (closed symbols) and powder (open symbols) samples of Yb$_2$Ti$_2$O$_7$ on a semi-logarithmic scale. The lines are a guide to the eye. Data from Ross \textit{et al.}~\cite{Ross:11}, Yaouanc \textit{et al.}~\cite{Yaouanc:11}, D'Ortenzio \textit{et al.}~\cite{DOrtenzio:13}, and Bl\"{o}te \textit{et al.}~\cite{Blote:69} are also shown for comparison. 
(b) Left-hand $y$-axis: magnetic entropy versus temperature $S(T)$. The phonon contribution, estimated from $C(T)$ data for non-magnetic Y$_2$Ti$_2$O$_7$ and corrected for the difference in the atomic mass of Y and Yb~\cite{Bouvier:91}, and a nuclear Schottky contribution~\cite{Blote:69}, have been subtracted from the data.
Right-hand $y$-axis: the flipping ratio of the neutron spins at the (111) Bragg peak as a function of temperature for the single crystal sample of Yb$_2$Ti$_2$O$_7$ used in this work (see Ref.~\onlinecite{Chang:12}).
% These data emphasize the first-order nature of the magnetic transition. 
The vertical black and red dashed lines indicate the temperature of the specific heat peak for the crystal and powder, respectively.
}
  \label{fig:C}
\end{figure}

Polycrystalline samples of Yb$_2$Ti$_2$O$_7$ were synthesized using a solid state reaction method. Stoichiometric quantities of Yb$_2$O$_3$ and TiO$_2$ powders were repeatedly ground, pressed into pellets, and sintered at 1300~$^\circ$C for several days. The method used for the growth of the single-crystal is described in Ref.~\onlinecite{Chang:12}. Heat capacity measurements were carried out using a relaxation method in a Quantum Design Physical Property Measurement System (PPMS) equipped with a $^3$He-$^4$He  dilution refrigerator insert. Figure~\ref{fig:C}(a) shows the specific heat data of this new powder sample and the single-crystal used in Refs.~\onlinecite{Chang:12} and ~\onlinecite{Yasui:03}, as well as data published by other groups. From the peak in the specific heat, the transition temperature $T_\mathrm{C}$ for the powder and the single crystal are estimated to be 0.26 and 0.21~K respectively. Note, the $T_\mathrm{C}$ for our new powder sample is higher than that of Ref.~\onlinecite{Blote:69} and as high as those of Refs.~\onlinecite{Yaouanc:11} and~\onlinecite{DOrtenzio:13}. For both samples, the magnetic entropy released between 50~mK and 30~K amounts to $R\ln2$ per Yb ion, indicating no residual entropy for the lowest-energy magnetic doublet of the Yb $4f$ electrons,\cite{Blote:69} as shown in Fig.~1(b). Although the relaxation method used here cannot correctly extract the latent heat, the first-order nature of the ferromagnetic transition at least in the case of our single crystal is clear from a thermal hysteresis of the flipping ratio of the neutron spins at the (111) Bragg peak, which decays to unity below $T_\mathrm{C}$~\cite{Chang:12} (see Fig.~1(b)). 
\subsection{$\mu$SR EXPERIMENTS}
The $\mu$SR measurements on our powder and single-crystal samples were performed using the MuSR spectrometer at ISIS where a pulse of muons with a full-width at half maximum (FWHM) of $\sim70$~ns is produced every 20~ms. The samples were mounted on pure Ag plates using General Electric (GE) varnish and covered with silver foil. The samples were cooled to 50~mK in an Oxford Instruments $^3$He-$^4$He dilution refrigerator and the data were collected on heating. The 100\% spin-polarized muons are implanted into a sample and after coming to rest the muon spin precesses in the local magnetic environment. The muons decay with a half-life of $2.2~\mu$s, emitting a positron preferentially in the direction of the muon spin at the time of decay. Two sets of detectors, up and down stream of the sample, count the number $F(t)$ and $B(t)$, of decay positrons emitted from the sample. The asymmetry of the $\mu$SR time spectrum is then obtained as $A(t)=(F(t)-\alpha B(t))/(F(t)+\alpha B(t))$, where $\alpha$ represents a relative counting efficiency of the forward and backward detectors.\cite{Lee:book, Yaouanc:book} The data were normalized after subtracting a well characterized constant background signal in the asymmetry time spectra that arises from the high-purity silver sample holder. For the ZF muon experiments, stray fields at the sample position were compensated to within $1~\mu$T by an active compensation coils system. LF spectra were collected with magnetic fields up to 0.25~T applied along the direction of the incident muons.

\section{EXPERIMENTAL RESULTS AND DISCUSSION}
\begin{figure}[tb]
  \begin{center}\leavevmode
    \includegraphics[angle=-90,width=0.8\columnwidth]{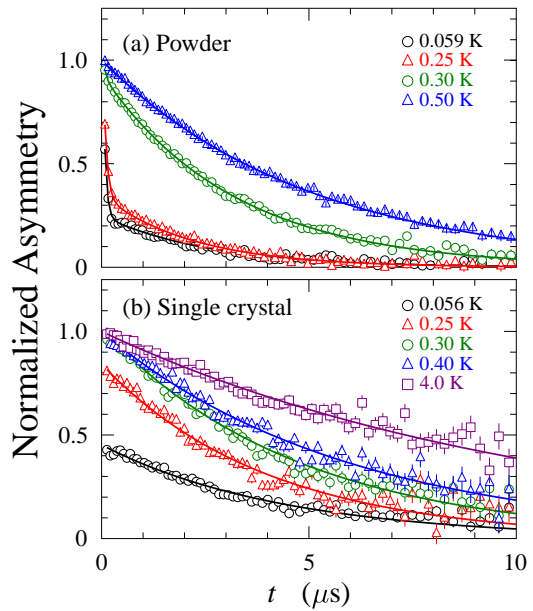}
  \end{center}
  \caption{(Color online). ZF $\mu$SR asymmetry time spectra for (a) the powder and (b) the single crystal samples of Yb$_2$Ti$_2$O$_7$ between 0.05 and 4~K.}
  \label{fig:ZFmuSR}
\end{figure}

Let us start with the temperature dependence of the ZF asymmetry time spectra, which are shown in Fig.~\ref{fig:ZFmuSR} for the powder and the single crystal. Both data sets have been normalized by the value $A(t=0,\ T=0.6~\mathrm{K})$.
In the case of the powder sample, the asymmetry shows an exponential decay with time at 0.5~K. This indicates that the Yb spins are totally fluctuating in time, yielding a slow relaxation of the muon spins. The relaxation becomes faster with decreasing temperature down to 0.3~K. Eventually, at 0.25~K, slightly below the $T_\mathrm{C}$ determined from the $C(T)$ data, the asymmetry exhibits a kink, which is characterized by a steep initial drop in a fast time domain within 0.5~$\mu$s, followed by a slow relaxation. The observation of an initial drop in the asymmetry indicates that the muon spins are depolarized more quickly than in the pulse duration of 70~ns.
This provides strong evidence that the sample contains magnetic moments which are either static or quasi-static within the $\mu$SR time window 10~ps to 1~$\mu$s.\cite{REF-ADD-1} Our results on this powder sample are comparable with previous results on other polycrystalline samples.\cite{Hodges:02} The single crystal shows similar behavior to our powder sample, but the relaxation of the muon spins is slightly slower at both 0.4 and 0.3~K. At $T\leq 0.25$~K, we again observe a rapid initial drop in the asymmetry followed by a slow relaxation, although the size of the initial decrease in asymmetry is reduced from that of the powder.

\begin{figure}[tb]
  \begin{center}\leavevmode
    \includegraphics[angle=-90,width=0.8\columnwidth]{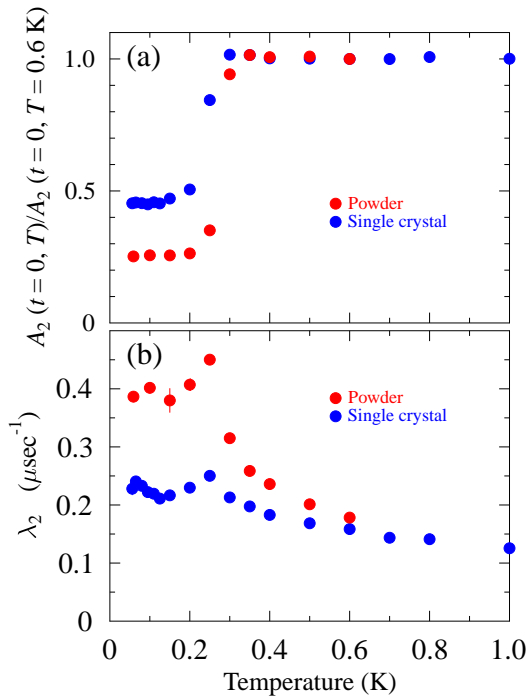}
  \end{center}
  \caption{(Color online). Temperature dependence of the normalized ZF $\mu$SR asymmetry, $A_2(t=0, T)/A_2(t=0, T=0.6~\mathrm{K})$, and the depolarization rate, $\lambda_2$, of the slowly depolarizing component for (a) powder and (b) single-crystal samples of Yb$_2$Ti$_2$O$_7$.}
  \label{fig:muSR_AD}
\end{figure}

These $\mu$SR time spectra can be analyzed successfully by using a two-exponential relaxation function,

\begin{equation}
\label{ZFfit}
A(t)=A_1e^{-\lambda_1t}+A_2e^{-\lambda_2t},
\end{equation}
\noindent with the first and second terms representing the fast and slow relaxation components, respectively, i.e., the relaxation rate $\lambda_1>\lambda_2$. This gives a minimal effective model for treating magnetic transitions and a magnetically ordered volume fraction in ordered phases.\cite{REF-ADD-1} Figure~\ref{fig:muSR_AD}(a) shows the value of the initial asymmetry, $A_2(t=0)$, normalized by the value at 0.6~K, for the slowly relaxing component of both the powder and single-crystal samples. $A_2(t=0)$ is nearly temperature independent at high temperatures, starts to decrease below 0.35 and 0.3~K for the powder and the single crystal respectively, and falls to an almost constant value below 0.2~K. Note that $A_2(t=0)$ reduces to about $\tfrac{1}{4}$ and  $\tfrac{1}{2}$ of its high-temperature values for the powder and the single crystal, respectively. This means that the volume fraction of static or quasi-static magnetic moments is estimated to be nearly $100\%$ for the powder, since it is close to an ideal value  $\tfrac{1}{3}$, and about $80\%$ for the single crystal.\cite{REF-ADD-1}
Figure~\ref{fig:muSR_AD}(b) shows the relaxation rate $\lambda_2$ for the slowly relaxing component of both the powder and single-crystal samples. The relaxation rates are almost the same at around 1~K, are gradually enhanced with decreasing temperature, reach the maximum around 0.25~K, and are nearly saturated at lower temperatures. However, $\lambda_2$ for the powder data exhibits a stronger enhancement by a factor of about two, as compared with the single-crystal data. The peak of $\lambda_2$ around 0.25~K indicates that the time scale of magnetic fluctuations is slowing down and crosses the $\mu$SR time window, 10~ps - 1~$\mu$s, around this temperature. Although it is thought that the transition here is first-order, it is accompanied by a moderately high intensity of slow magnetic fluctuations.

\begin{figure}[tb]
  \begin{center}\leavevmode
    \includegraphics[angle=-90,width=\columnwidth]{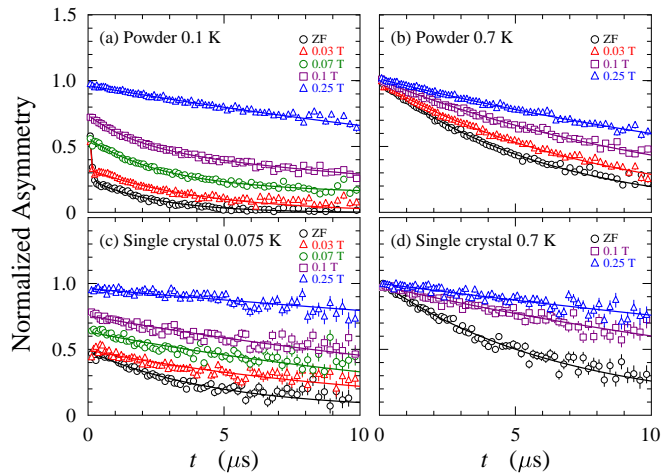}
  \end{center}
  \caption{(Color online). LF $\mu$SR asymmetry time spectra for polycrystalline Yb$_2$Ti$_2$O$_7$ at (a) 0.1 and (b) 0.7~K and single-crystal Yb$_2$Ti$_2$O$_7$ at (c) 0.075 and (d) 0.7~K.}
  \label{fig:LFmuSR}
\end{figure}

To confirm that the ``nearly static magnetic moments'' observed
with the ZF $\mu$SR results are indeed static, we have also
measured the asymmetry with longitudinal fields applied along the direction of the
initial muon spin polarization. Figures~\ref{fig:LFmuSR}(a)
and~\ref{fig:LFmuSR}(b) show the LF $\mu$SR time spectra observed
at 0.1 and 0.7~K on the powder sample and Figs.~\ref{fig:LFmuSR}(c)
and \ref{fig:LFmuSR}(d) those at 0.075 and 0.7~K for the single
crystal. Again, all the curves are normalized by the value $A(t=0, T=0.6~\mathrm{K})$ at zero field and then fit using Eq.~\eqref{ZFfit}. Clearly, the slower relaxation rate
$\lambda_2$ decreases monotonously with increasing longitudinal
field both well below and well above $T_\mathrm{C}$. In particular, at
0.7~K, the asymmetry still shows a single slow exponential time
decay even in a magnetic field of 0.25~T. This indicates that all the Yb spins are
fluctuating and that some of the magnetic excitation spectra lie in
the $\mu$SR time window in this field range. On the other hand,
well below $T_\mathrm{C}$ (at 0.1~K for the powder and at 0.075~K for the
single-crystal), with increasing field, the whole asymmetry spectra
shows an upward parallel shift with no crossing. This shift is
accompanied by a recovery of the initial loss in the asymmetry and
a reduction of the long-time depolarization rate $\lambda_2$. This
decoupling behavior seen in the time spectrum is typical for a case
in which a static internal magnetic field appears at a muon site.
Namely, this provides evidence that the magnetic moments inside both the powder and single-crystal samples are indeed static. Note that an observation of the muon spin precession is not necessary to draw this conclusion.\cite{REF-ADD-2} 

We can try to estimate the magnitude of the ordered internal magnetic field. To do this we follow a method used previously~\cite{REF-ADD-2} and discussed in detail in Ref.~\onlinecite{REF-ADD-4}, The equation adopted is

\begin{equation}
\label{internalH}
A_2 = A_0 \left \{\frac{3}{4}-\frac{1}{4x^2}+\frac{(x^2-1)^2}{16x^3}\ln \frac{(x+1)^2}{(x-1)^2} \right \},
\end{equation}
\noindent where $x= H_{\text{LF}} / H_{\text{int}}$, $H_{\text{LF}}$ is the applied external field, and $H_{\text{int}}$ is the internal field at the muon site. There are certain conditions for this function to be applicable. The internal fields at muon sites must have the same amplitude but their directions at different sites throughout the sample may be random. These conditions can be realized in powder samples. Adopting this method the magnitude of the ordered internal magnetic field for the polycrystalline sample can be estimated from the initial drop in the asymmetry at zero field to be $65\pm3$~mT. A similar calculation for the single-crystal sample gives a value of $71\pm4$~mT. This latter value must be treated with some caution given the limitations of the analysis discussed above. Nevertheless, we included it here for completeness. 
We also note that the initial drop in the asymmetry eventually disappears at 0.25~T, which is actually of the order of $T_\mathrm{C}$.

We now compare our results with previous work. Hodges \textit{et al.}~\cite{Hodges:02} reported the results of ZF $\mu$SR measurements on a powder sample. Their results are in reasonable agreement with ours, but are insufficient to conclude that the magnetic moments in the sample are either static or quasi-static, because they did not perform LF $\mu$SR measurements. Our LF $\mu$SR data for the powder sample are quite similar to those obtained by de R\'eotier \textit{et al.}, although the presence of a clear minimum in their relaxation data at short times led them to suggest that the muons were detecting a field distribution influenced by disorder, and that this disorder probably resulted from the geometrical frustration of the magnetic interactions.\cite{REF-ADD-3} They used a Gaussian-broadened Gaussian function that assumed that the muons see different environments, each of them characterized by a dynamical Kubo-Toyabe function, to fit their data and concluded that Yb$_2$Ti$_2$O$_7$ presents dynamical short-range correlations at low-temperature. There is no strong evidence for a shallow minimum at short times in the LF relaxation data for our powder sample, although there is a hint of such a dip in the data collected in 0.03~T (see Fig.~\ref{fig:LFmuSR}a). We see no dip the LF asymmetry spectra for the single-crystal. As discussed above, our data are consistent with the moments being static, while a disordered but static scenario is precluded by the absence of any residual magnetic entropy at the lowest temperature in our heat capacity data.

Recently, D'Ortenzio \textit{et al.} also performed ZF and transverse-field $\mu$SR experiments on different powder and single-crystal samples.\cite{DOrtenzio:13} However, their asymmetry time spectra never show any significant difference in the initial ($t=0$) value between at 1~K and 16~mK, which are strikingly different from those obtained by Hodges \textit{et al.}\cite{Hodges:02} and in our work. 
%However, our current ZF and LF $\mu$SR experiments on the powder as well as on the single crystal clearly evidences that the magnetic moments are static below $T_\mathrm{C}$. Note that it is usually quite hard to detect a small ($6\%$~\cite{Yasui:03}) increase of the Bragg peak intensity due to magnetic contribution in neutron diffraction experiments on powder samples.

The differences between our $\mu$SR results and those of Ref.~\onlinecite{DOrtenzio:13} could be ascribed to a sample dependence issue. In particular, the specific heat of single crystals strongly depends on the sample. For the single crystals studied in Refs.~\onlinecite{Yaouanc:11} and ~\onlinecite{DOrtenzio:13}, $C(T)$ only exhibits a broad bump followed by a large tail continuing down to 0.1~K, at which temperature the value of $C(T)$ is about five times larger than that of our powder sample (see Fig.~\ref{fig:C}). The single-crystal $C(T)$ data of Ref.~\onlinecite{Ross:12}, in which long-range magnetic order was not observed with neutron scattering,\cite{Ross:11} has a peak at 0.26 and broad shoulder around 0.20~K. To the best of our knowledge, $\mu$SR experiments have not been reported on this sample. Our single-crystal exhibits a single sharp peak in $C(T)$ at $\sim0.21$~K. The magnitude of $C(T)$ at 0.1~K is still double that of our powder sample. This is qualitatively consistent with the estimated $80\%$ for the static magnetic volume fraction of our single-crystal sample compared with an almost perfect ($100\%$) magnetic volume fraction for our powder sample.

In the case of powder samples, the differences between our $\mu$SR results and those of Ref.~\onlinecite{DOrtenzio:13} cannot simply be understood from the behavior of the specific heat, since the different powder samples used for our work, by Yaouanc \textit{et al.},\cite{Yaouanc:11}, and by D'Ortenzio \textit{et al.},\cite{DOrtenzio:13} all exhibit similar anomalies in $C(T)$ at the temperature of $\sim0.26$~K which is higher than the $C(T)$ anomaly reported in the early powder work~\cite{Blote:69} as well as for our single crystal, see Fig.~\ref{fig:C}. Finally we note that in the case of the first-order phase transition expected here, measurements of $C(T)$  made using a heat pulse-relaxation method may not fully determine the true peak height, since one needs to measure the latent heat.

\section{CONCLUSIONS}
Our results provide compelling evidence that in both our powder and single-crystal samples of Yb$_2$Ti$_2$O$_7$, the Yb moments undergo long-range magnetic order below $T_\mathrm{C}$ ($\sim0.26$~K for the powder and $\sim0.21$~K for the single crystal). The $\mu$SR results indicate the formation of static magnetic moments, while the loss of magnetic entropy from high temperature to well below $T_\mathrm{C}$ indicates that these static magnetic moments are ordered. 

The data are consistent with our previous observations on the same single-crystal sample of a ferromagnetic order with an extremely slow relaxation by neutron diffraction~\cite{Yasui:03} and polarized diffuse neutron scattering experiments~\cite{Chang:12}. The suppression of the low-frequency magnetic excitation spectral density within the $\mu$SR time window below $T_\mathrm{C}$ is compatible with a Higgs mechanism by which pyrochlore ``photons'' acquire an energy gap due to a coupling to spinons that carry emergent magnetic monopoles.~\cite{Hermele:04}

Uncertainties in estimating both the magnetically static volume fraction using $\mu$SR and the magnetic entropy associated with order from heat capacity data, leave open the possibility that either a fraction of the magnetic moments throughout the bulk of the sample, or all the moments in some macroscopic portion of the samples (especially for the single-crystal), remain dynamic or disordered. Using the data presented here, it is problematic to comment on the structure of the magnetic order, i.e. whether it is a simple ferromagnet or has a ferromagnetic component. Why some samples, and particularly some single-crystals, appear to exhibit long-range magnetic order while others do not also remains an open issue. Clearly more research, and especially work which correlates the low-temperature magnetic properties with the structure and chemistry of Yb$_2$Ti$_2$O$_7$ will be required, before the properties of this fascinating and important material are fully understood.

\begin{acknowledgments}
The authors thank Geetha Balakrishnan and Ravi Singh for help with
sample preparation, Yixi Su for fruitful discussions, and the STFC for beam time. This work is partially
supported by the National Science Council, Taiwan under grant no.
NSC 101-2112-M-006-010-MY3 (L.J.C.), by Japan Society of the Promotion of Science under grant no. 24740253 from (S.O.), by
the EPSRC, United Kingdom grant no. EP/I007210/1, and by the Royal
Society, United Kingdom by the International Exchanges Scheme grant
no. IE110911 (L.J.C. and M.R.L). Some of the equipment used in this
research was obtained through the Science City Advanced Materials
project: Creating and Characterizing Next Generation Advanced
Materials project, with support from Advantage West Midlands (AWM)
and part funded by the European Regional Development Fund (ERDF).
\end{acknowledgments}


\begin{thebibliography}{99}

\bibitem{Hermele:04}
M. Hermele, M. P. A. Fisher, and L. Balents,
%Ref1
%Pyrochlore photons: the $U(1)$ spin liquid in a S=$\frac{1}{2}$ three-dimensional frustrated magnet.
Phys. Rev. B \textbf{69}, 064404 (2004).

\bibitem{Gardner:10}
J. S. Gardner, M .J. P. Gingras, and J. E. Greedan,
%Ref2 Magnetic pyrochlore oxides.
Rev. Mod. Phys. \textbf{82}, 53 (2010).

\bibitem{Chang:12}
%Ref3
L.-J. Chang, S. Onoda, Y. Su, Y.-J. Kao, K.-D. Tsuei, Y. Yasui, K. Kakurai, and M. R. Lees,
Nat. Commun. \textbf{3}, 992 (2012).

\bibitem{Anderson:63}
P. W. Anderson,
%Ref4 Plasmons, Gauge Invariance, and Mass.
Phys. Rev. \textbf{130}, 439 (1963).

\bibitem{Higgs:64}
P. W. Higgs, Phys. Rev. Lett. \textbf{13}, 508 (1964).

\bibitem{Fradkin:79}
E. Fradkin, and S. H. Shenker,
%Phase diagrams of lattice gauge theories with Higgs fields.
Phys. Rev. D \textbf{19}, 3682 (1979)

\bibitem{Blote:69}
W. J. Bl\"{o}te, R. F. Wielinga, and W. J. Huiskamp,
%Heat-capacity measurements on rare-earth double oxides R$_2$M$_2$O$_7$.
Physica (Amsterdam) \textbf{43}, 549 (1969).

\bibitem{Hodges:02}
J. A. Hodges, P. Bonville, A. Forget, A. Yaouanc, P. Dalmas de R\'eotier, G. Andr\'e, M. Rams, K. and Kr\'olas, C. Ritter, P. C. M. Gubbens, C. T. Kaiser, P. J. C. King, and C. Baines,
%First-order transition in the spin dynamics of geometrically frustrated Yb$_2$Ti$_2$O$_7$.
Phys. Rev. Lett. \textbf{88}, 077204 (2002).

\bibitem{Yasui:03}
Y. Yasui, M. Soda, S. Iikubo, M. Ito, M. Sato, N. Hamaguchi, T. Matsushita, N. Wada, T. Takeuchi, N. Aso, and K. Kakurai,
%Ferromagnetic transition of pyrochlore compound Yb$_2$Ti$_2$O$_7$.
J. Phys. Soc. Jpn. \textbf{72}, 3014 (2003).


\bibitem{Ross:11}
K. A. Ross, L. Savary, B. D. Gaulin, and L. Balents,
%Quantum excitations in quantum spin ice.
Phys. Rev. X \textbf{1}, 021002 (2011).

\bibitem{Yaouanc:11}
A. Yaouanc, P. Dalmas de R\'eotier, C. Marin, and V. Glazkov,
%Single-crystal versus polycrystalline samples of magnetically frustrated Yb$_2$Ti$_2$O$_7$: specific heat results.
Phys. Rev. B \textbf{84}, 172408 (2011).

\bibitem{Ross:12}
K. A. Ross, T. Proffen, H. A. Dabkowska, J. A. Quilliam, L. R. Yaraskavitch, J. B. Kycia, and B. D. Gaulin,
%  Lightly stuffed pyrochlore structure of single-crystalline Yb2Ti2O7 grown by the optical floating zone technique
Phys. Rev. B \textbf{86}, 174424 (2012).

\bibitem{DOrtenzio:13}
R. M. D'Ortenzio, H. A. Dabkowska, S. R. Dunsiger, B. D. Gaulin, M. J. P. Gingras, T. Goko, J. B. Kycia, L. Liu, T. Medina, T. J. Munsie, D. Pomaranksi, K. A. Ross, Y. J. Uemura, T. J. Williams, and G. M. Luke,
Phys. Rev. B \textbf{88}, 134428 (2013).

\bibitem{Henley:05}
C. L. Henley,
%Power-law spin correlations in pyrochlore antiferromagnets.
Phys. Rev. B \textbf{71}, 014424 (2005)

\bibitem{Onoda:10}
S. Onoda and Y. Tanaka,
%Quantum melting of spin ice: emergent cooperative quadrupole and chirality.
Phys. Rev. Lett. \textbf{105}, 047201 (2010);
%Quantum fluctuations in the effective pseudospin-$\frac{1}{2}$ model for magnetic pyrochlore oxides.
Phys. Rev. B \textbf{83}, 094411 (2011);
S. Onoda,
%Effective quantum pseudospin-$1/2$ model for Yb pyrochlore oxides.
J. Phys.: Conf. Series. \textbf{320}, 012065 (2011).

%\bibitem{Banerjee:08}
%A. Banerjee, S. V. Isakov, K. Damle, and Y. B. Kim,
%Unusual liquid state of hard-core bosons on the pyrochlore lattice
%Phys. Rev. Lett. \textbf{100}, 047208 (2008).

\bibitem{Gardner:04}
J. S. Gardner, G. Ehlers, N. Rosov, R. W. Erwin, and C. Petrovic,
%Spin-spin correlations in Yb$_2$Ti$_2$O$_7$: a polarized neutron scattering study.
Phys. Rev. B \textbf{70}, 180404(R) (2004).

\bibitem{Bouvier:91}
M. Bouvier, P. Lethuillier, and D. Schmitt,
%Specific heat in some gadolinium compounds. I. Experimental,
Phys. Rev. B \textbf{43}, 13137 (1991).

\bibitem{Lee:book}
S. L. Lee, S. H. Kilcoyne, and R. Cywinski, \textit{Muon Science:Muons in Physics, Chemistry and Materials} (SUSSP Publications and IOP Publishing, Bristol, 1999).

\bibitem{Yaouanc:book}
A. Yaouanc and P. D. de R\'eotier, \textit{Muon Spin Rotation, Relaxation, and Resonance} (Oxford University Press, New York, 2011).


\bibitem{REF-ADD-1}
T. Adachi, S. Yairi, K. Takahshi, Y. Koike, I. Watanabe, and K. Nagamine, Phys. Rev. B \textbf{69}, 184507 (2004).


\bibitem{REF-ADD-2}
H. Hachitani, H. Fukazawa, Y. Kohori, I. Watanabe, C. Sekine, and I. Shirotani, Phys. Rev. B \textbf{73}, 052408 (2006).


\bibitem{REF-ADD-3}
P. Dalmas de R\'eotier, V. Glazkov, C. Marin, A. Yaouanc, P. C. M. Gubbens, S. Sakarya, P. Bonville, A. Amato, C. Baines, P. J. C. King, Physica B \textbf{374-375}, 145 (2006).

\bibitem{REF-ADD-4}
F. L. Pratt, J. Phys.: Condens. Matter \textbf{19}, 456207 (2007).


\end{thebibliography}
\end{document}